\documentstyle[prb,aps,twocolumn,epsf]{revtex}
\input{epsf}
\newcommand{\LL}[1]{\marginpar{\protect\tiny \protect\setlength
   {\baselineskip}{8pt} !!!}}
\newcommand{\AL}[2]{\marginpar{\protect\tiny \protect\setlength
   {\baselineskip}{8pt} ?? }}
\begin{document}
\draft

\twocolumn[\hsize\textwidth\columnwidth\hsize\csname @twocolumnfalse\endcsname

\title{\bf Structural properties of silicon dioxide thin films densified by
medium-energy particles}

\author{
Alexis Lef\`evre,$^a$ Laurent J.\ Lewis,$^{a}$\cite{byline2} Ludvik
Martinu,$^b$ Michael R.\ Wertheimer$^b$
}

\address{
$^{a}$D\'epartement de Physique et Groupe de Recherche en Physique et
Technologie des Couches Minces (GCM), Universit\'e de Montr\'eal, Case
Postale 6128, Succursale Centre-Ville, Montr\'eal, Qu\'ebec, Canada H3C 3J7
}
\address{
$^{b}$D{\'e}partement de G\'enie Physique et de G\'enie des Mat\'eriaux et
Groupe de Recherche en Physique et Technologie des Couches Minces (GCM),
\'Ecole Polytechnique de Montr\'eal, Case Postale 6079, Succursale
Centre-Ville, Montr\'eal, Qu\'ebec, Canada H3C 3A7
}

\date{\today}

\maketitle

\begin{center}
Submitted to {\em Physical Review B}
\end{center}

\begin{abstract}

Classical molecular-dynamics simulations have been carried out to investigate
densification mechanisms in silicon dioxide thin films deposited on an
amorphous silica surface, according to a simplified ion-beam assisted
deposition (IBAD) scenario. We compare the structures resulting from the
deposition of near-thermal (1 eV) SiO$_{2}$ particles to those obtained with
increasing fraction of 30 eV SiO$_{2}$ particles. Our results show that there
is an energy interval --- between 12 and 15 eV per condensing SiO$_2$ unit on
average --- for which the growth leads to a dense, low-stress amorphous
structure, in satisfactory agreement with the results of low-energy ion-beam
experiments. We also find that the crossover between low- and high-density
films is associated with a tensile to compressive stress transition, and a
simultaneous healing of structural defects of the {\em a-}SiO$_2$ network,
namely three- and four-fold rings. It is observed, finally, that densification
proceeds through significant changes at intermediate length scales (4--10 \AA),
leaving essentially unchanged the ``building blocks'' of the network, viz.\ the
Si(O$_{1/2}$)$_{4}$ tetrahedra. This latter result is in qualitative agreement
with the mechanism proposed to explain the irreversible densification of
amorphous silica recovered from high pressures ($\sim$ 15--20 GPa).

\end{abstract}


\maketitle

\vskip2pc]

\narrowtext

\section{Introduction}

In response to the rapidly increasing demand for highly-specialized
applications in optical filtering devices, integrated electronics and
photonics technology, considerable efforts have been devoted in the past few
decades to the development of methods for growing thin silicon dioxide
(SiO$_2$) coatings having low stress and low defect concentrations, and
exhibiting bulk-like optical properties. To this end, methods commonly
referred to as ion-assisted deposition (IAD), in which growth is accompanied
by a flux of energetic particles, are of particular
interest:\cite{poitrasrev,pulker,ross,smidt} indeed, bombardment with low to
medium-energy neutral or ionic particles --- in the range 10--200 eV --- is
found to have beneficial effects on a number of physical properties of the
deposited material. Besides fulfilling the above requirements, it also leads
to reduced moisture absorption, higher density, lower chemical etch rate,
better adhesion to the substrate, and therefore better overall long-term
behavior.\cite{brunet,bazylenko,souche} In addition, IAD methods allow
processing at lower substrate temperatures, making them especially adapted to
growth on polymers or graded-index coatings.\cite{poitrasrev} While there
evidently exists a threshold energy below which benefits are insignificant,
too high an energy can have undesirable consequences, such as preferential
sputtering or impurity incorporation.\cite{martinu}

In ion-beam assisted deposition (IBAD) --- a special case of IAD --- an
external source provides a well collimated flux of ions with a narrow kinetic
energy distribution. It has proven useful to discuss the dependance of
structural changes of IBAD films upon bombardment in terms of three
parameters: the mean kinetic energy of the ions, $E_i$, the ratio of
ion-to-condensing particle fluxes at the surface, $\phi_i/\phi_n$ (where
$\phi_n$ is measured from the growth rate), and the energy delivered by each
particle that condenses, $E_p=E_i \phi_i/\phi_n$.\cite{ross,brunet,souche}
It has been observed that there exist ``critical'' values of these parameters
below which the films are porous and above which they are
densely packed.\cite{ross} In the latter case, films are in general found
to exhibit better physical properties, closer to their bulk, relaxed
counterparts. For SiO$_2$, $E_p$, determined from optical index, step
coverage or optical transmittance measurements, is in the range 10--100 eV,
depending on the particular set of parameters used in the growth
process.\cite{ross,brunet,souche,mcneil}

Clearly, bombardment profoundly affects the microstructure of the deposited
material. For amorphous solids --- which is the case of SiO$_2$ films ---
static disorder hampers the measurement of detailed information beyond the
second-neighbor shell. As a consequence, experimental results must be
considered in the light of structural models. In this way, proper conclusions
on the atomic mechanisms that mediate bombardment-induced densification can
be drawn. Thus, for instance, models help understanding how changes in the
medium-range structure of amorphous silica ({\em{}a-}SiO$_2$) influence the
Si--O--Si bond-angle,\cite{vash7} which in turn is related to such properties
as the density, the optical index or the amount of built-in
stress.\cite{poitrasrev,brunet,souche,lucovsky}

It is useful to discuss the structure of {\em a-}SiO$_2$ in terms of
short-range order (SRO) and medium-range order (MRO)
correlations,\cite{vash3} at length scales $\le$ 4 \AA\ and in the range 4--10
\AA, respectively. SRO is well defined in terms of such quantities as bond
lengths and coordination numbers. It is mainly associated to the structural
``building block'' of the network, the Si(O$_{1/2}$)$_{4}$ tetrahedron. Both
numerical simulations\cite{vash0,tse,stixrude} and
experiment\cite{devinesiosi,devinesio,susman} have shown the SRO correlations
in permanently-densified (20\%) {\em a}-SiO$_2$ recovered from high pressures
($\sim$ 15-20 GPa) to be essentially the same as in the ``normal'' material.

In contrast, MRO correlations, which are at the origin of the so-called
``first sharp diffraction peak'' (FSDP), undergo important modifications upon
compression:\cite{susman,tan,sugai} the height of the FSDP decreases and its
position increases with density. Molecular-dynamics\cite{vash0} and
Monte-Carlo\cite{stixrude} simulations have attributed this effect to
modifications in the topology of the network; in particular, the intensity of
the FSDP is reported to be roughly proportional to the number of six-membered
rings. (A $n$-membered ring is defined as a closed loop of $n$ Si--O bonds).
Further, the x-ray diffraction spectra of porous silica thin films deposited
at low substrate temperatures, using either electron-beam
evaporation\cite{himmel} or low-pressure chemical vapor deposition
(LPCVD),\cite{konnert} exhibit a strong low-angle scattering intensity and a
FSDP height smaller than that of bulk {\em a}-SiO$_2$. The latter effect,
which diminishes upon annealing, has been attributed to differences at the MRO
level. However, to the best of our knowledge, no systematic studies of the
changes in MRO with the kinetic energy of the ions or the ion-current density
have so far been reported; this constitutes one objective of the present work.

Several computational methods have been employed to assess the role of ion
beams during thin film growth.\cite{revmodtfg} Among them, molecular-dynamics
(MD) simulations have provided useful insights into the understanding of
ion-induced layer-by-layer growth,\cite{jacobsen} surface mobility,\cite{ohira}
pore annihilation,\cite{dong} stress\cite{fang} and defect
formation.\cite{kitabatake} These studies have focused on the role of
collision-induced events for obtaining crystalline material. Although the
precise values of the parameters required for this purpose are
material-dependent, a beam energy of the order of a few tens of eV --- in
agreement with experiment\cite{jacobsen,hubler} --- appears to be adequate,
yielding ordered structures on the scale of many interatomic spacings. These
simulations allow collision-induced events to be sorted out and relevant
deposition parameters to be identified, thus providing useful information for
better control of the growth process.

The situation is more complicated for the case of amorphous materials growth,
as spatial order is short-ranged. Only very few MD studies of the structural
properties of amorphous thin film growth have thus far been reported; notable
exceptions are Refs.\ \onlinecite{strickland,hensel} for silicon and Refs.\
\onlinecite{kaukonen,jager} for diamond-like carbon films. In the latter
case, bonding is found to be partly $sp^2$ and partly $sp^3$, that is,
between graphite and diamond. For the case of SiO$_2$ --- for which no such
studies so far exist --- considering the large number of polymorphs of the
material,\cite{liebau} as-deposited films are expected to show a large
diversity of structures as a function of density.

In view of this, and given the technological importance of the material,
we have carried out detailed MD simulations of the growth of silicon
dioxide on {\em a-}SiO$_2$ substrates. More precisely, growth is modelled
within a simplified IBAD scenario: we examine, at the atomic level, how
the structure of the films evolves as a function of $R$, the ratio of
medium- (30 eV) to low-energy (1 eV) SiO$_{2}$ particle fluxes impinging
on the surface. The rationale for the model is given in Sec.\ \ref{proc}.
For the sake of simplicity, the energetic ions are taken to be SiO$_{2}$
``molecules''; although this is a very crude approximation to the real
growth process, our model is expected to provide much-needed information
into the fundamental aspects of bombardment-induced densification. We
find, on average, that densification is related to changes in the ring
structure, in a way which is similar to the case of pressure-induced
densification of ``normal'' (2.2 g/cm$^3$) silica glass. The intensity of
the FSDP and the number of six-membered rings are here found to increase
with density. We also find that the crossover from low- to high-density
films is associated with a tensile-compressive stress transition,
occurring for $E_p$ of about 12--15 eV. It is observed, finally, that
medium-energy bombardment promotes the annihilation of structural defects,
namely three- and four-membered rings, which are at the origin of the
$D_2$ and $D_1$ defect lines observed in the Raman spectrum of
silica.\cite{devine}

\section{Computational details}\label{proc}

MD consists in integrating the classical equations of motion for a set of
interacting particles. The interactions can be derived from first principles,
or expressed in some effective, empirical form whose parameters are fitted to
experimental data. The first-principles approach is very accurate, but does
not allow large systems to be dealt with over ``long'' times cales. These
limitations are not so much of a problem in the empirical approach, but
accuracy and ``transferability'', that is, the ability for a model potential
to adequately describe a phase to which it was not fitted, is an issue. For
the problem considered here, empirical potentials are unavoidable. The model
developed by Nakano {\em et al.} has been successfully employed to describe
the bulk and the surface of disordered phases of SiO$_2$, {\em viz.}\
amorphous and porous
silica,\cite{vash7,vash0,vash1,vash2,vash4,vash5,vash6,trioni} which are
evidently relevant to the present problem. In this model, silicon is assumed
to be coordinated to four oxygen atoms, that is, the network is chemically
ordered. The total potential energy of the system is the sum of two- and
three-body contributions:
   \begin{eqnarray}
       V = \sum_{i<j}{V^{(2)}_{ij}(r) } +
           \sum_{i,j<k}{V^{(3)}_{ijk}(\vec{r}_{ij},\vec{r}_{jk})},
   \end{eqnarray}
where $i$, $j$, and $k$ run over all particles. The two-body term is written
as
   \begin{eqnarray}
      V^{(2)}_{ij}(r) & = & A_{ij}(\frac{\sigma_{i}+\sigma_{j}}{r})^{\eta_{ij}} +
                            \frac{Z_{i}Z_{j}}{r}e^{-\frac{r}{\lambda}} \nonumber\\
                      &   & - \frac{\alpha_{j}Z_{i}^{2}+\alpha_{i}Z_{j}^{2}}
                      {2r^{4}}e^{-\frac{r}{\xi}},
   \end{eqnarray}
where the various terms represent the steric repulsion, the Coulomb
interaction due to charge transfer, and a charge-dipole interaction due to
atomic polarizability, respectively. The three-body term, which describes the
covalent bonding, is expressed as
   \begin{eqnarray}
      V^{(3)}_{ijk}(\vec{r}_{ij},\vec{r}_{jk}) & = & B_{i}
         \exp(\frac{\mu}{r_{ij}-r_{0}}+\frac{\mu}{r_{jk}-r_{0}}) \nonumber\\
         & & \times [\cos(\theta_{ijk})-\cos(\theta^{0}_{ijk})]^{2} \nonumber\\
         & & \times \Theta(r_{0}-r_{jk})\Theta(r_{0}-r_{ij}),
   \end{eqnarray}
where $\Theta$ is the Heaviside step function, and $\theta^{0}_{ijk}$ is the
equilibrium value of the $(\vec{r}_{ji},\vec{r}_{jk})$ bond angle. This term
is calculated for nearest-neighbor Si--O--Si and O--Si--O triplets only. The
model was fitted to various physical properties of bulk SiO$_2$, such as
cohesive energy, elastic constants and the phonon density of states; the
values of the parameters can be found in Ref.\ \onlinecite{vash4}.

As demonstrated by surface extended x-ray absorption fine structure and x-ray
photoelectron spectroscopy measurements, the atomic structure of silica coatings
strongly depends on the particular deposition process used:\cite{forty,ray} both
silicon and oxygen atoms can be found in different environments, allowing the
presence of homoatomic bonds. This is particularly true when sub-oxides are
formed. These effects are not considered in our model; thus, the role of
structural defects --- such as peroxide radicals and, more generally, homopolar
bonds --- on relaxation mechanisms are not taken into account. For the case of
(nearly) stoichiometric situations, experimental data on SiO$_2$ produced by
plasma enhanced chemical vapor deposition (PECVD),\cite{bazylenko} RF
sputtering,\cite{forty} secondary ion deposition,\cite{george} and ion-beam
deposition\cite{albayati} indicate that the basic ``building-block'' of the
network is the Si(O$_{1/2}$)$_{4}$ tetrahedron. Thus, chemical ordering is
expected to dominate in these cases, and the model potential of Nakano {\em et
al.} is assumed to correctly describe atomic correlations.

In IBAD, silicon (di-)oxide is evaporated in a low-pressure oxygen
atmosphere;\cite{pulker,mcneil} part of the condensing particles therefore
have an energy smaller than 0.1 eV. Simultaneously, a beam of O$^+$/O$^+_2$
ions with energy in the range 30--500 eV is directed onto the substrate, in
order to promote surface relaxation as well as to hamper the formation of
suboxides due to preferential sputtering of oxygen atoms. Although the
experimental procedure is relatively simple, an atomistic description of it
is a formidable task, as it requires an environment-dependent potential for
oxygen ions, neutral gaseous particles, and for surface and bulk atoms. For
simplicity, because large systems have to be dealt with, we describe the
growth of silica in terms of a single type of condensing particle, namely
SiO$_{2}$, for {\em both} the deposited species and the incident energetic
ions. Since the clusters are free to dissociate on the substrate (and
actually do so), we expect this approximation to be of little consequence on
the final structure of the deposited film. However, because oxygen atoms are
preferentially sputtered as the deposited energy is raised (leading to a 10\%
deficit in the worst cases), SiO$_{3}$ clusters were introduced when
necessary so as to maintain the stoichiometry at 2:1. Also, the energy
transferred during a collision varies with the mass of the incoming particle,
which, in turn, may influence the value for which dense films are obtained.
Thus, care must be taken when comparing these results with experiment. We
also neglect atomic ionisation and electronic loss during collisions, a
reasonable approximation at low energies, especially for the case of
insulators.\cite{kaukonen,hedstrom} The present model is evidently crude, but
we expect that the generic features of the structure will be relevant to real
materials, even though the dynamics of the deposition process is not
described precisely.

The substrate on which growth takes place consists of a
$27.59\times27.59\times10$ \AA$^{3}$ amorphous slab containing 747 atoms,
extracted from a bulk ($27.59^3$ \AA$^{3}$) sample prepared using the
activation-relaxation technique (ART) of Barkema and Mousseau.\cite{mousseau}
The substrate is first subjected to an extensive thermal annealing cycle in
order to allow the surface to relax. A 3 \AA -thick layer at the bottom is
held fixed, in order to mimic a semi-infinite system, but also to prevent the
substrate from moving because of momentum transfer from incoming atoms. The
remaining 7 \AA\ are coupled to a heat bath at room temperature, consistent
with experiment.\cite{brunet,mcneil} Velocity renormalization was used to
maintain the system at constant temperature.

SiO$_{2}$ particles of roughly thermal kinetic energy ($E=0.02-0.06$ eV) were
used in an attempt to first grow samples without bombardment. However, this led
to unstable structures and to clusters of a dozen atoms desorbing from the
growing film. Much better results were obtained with $E=1$ eV, an energy
similar to that used by Hensel {\em et al.}\cite{hensel} (2 eV) in their MD
study of bombardment-induced epitaxial growth of elemental silicon. For the
high-energy particles, we used 30 eV, as suggested by the experimental results
of McNeil {\em et al.}\cite{mcneil} As noted previously, important process
parameters are expressed in term of the ion-to-condensing particle fluxes
ratio, $\phi_i/\phi_n$. However, this quantity is not known {\em a priori}
($\phi_n$ is deduced from the measurement of the growth rate); thus, it is
not a adequate parameter for monitoring the simulation procedure. Instead, we
used, as other authors,\cite{hensel} the ratio, $R$, of low-to-medium energy
particle fluxes. Low- and high-energy particles were chosen at random according
to the desired $R$ value, and deposited at normal incidence onto the substrate.
The equations of motion were integrated using the Verlet algorithm with an
adaptative time step, so as to adjust to abrupt changes of the velocities
during collisions.\cite{smith} The time interval between the deposition of two
successive particles was chosen in such a way as to provide a reasonable
compromise between computational effort and proper relaxation of the surface.
In practice, we found it adequate to relax the system during 4 and 8 ps after
depositing a low- and a high-energy particle, respectively, before a new
particle was introduced.

As the film grows, ``normal'' thermal diffusion into the substrate is not
sufficient to drain away the excess energy brought about by incoming
particles. Therefore, the entire sample was submitted to a 1 ps
thermalisation phase each time before introducing a new particle. Prior to
this, the temperature was around 350 K and 320 K for 30 eV and 1 eV SiO$_2$
particles, respectively. Since the constants of diffusion are small at
these temperatures, the equilibration procedure is expected not to
interfere with the growth process in any significant manner.

Simulations were carried out for several values of the parameter $R$, namely 0,
0.08, 0.22, 0.38, 0.55, 0.73, 1.07, 1.53, 2 and 2.73. It can be determined {\em
a posteriori} that these correspond to $\phi_i/\phi_n$ values of 0, 0.07, 0.18,
0.32, 0.41, 0.49, 0.62, 0.75, 0.80 and 0.9, respectively. In each case, about
1300 atoms ($\sim$430 clusters) were deposited. Following deposition, the
samples were equilibrated during 20 ps at 300 K, after which a 40-ps
microcanonical run was carried out in order to let the sample reach structural
equilibrium. The corresponding ``true'' duration of simulated growth is about 3
ns. All calculations were performed using the program {\tt groF}, a
general-purpose MD code for bulk and surfaces developed by one of the authors
(LJL).

\section{Results and discussion}\label{results}

\subsection{Structure of samples grown {\em without} bombardment}\label{without}

\begin{figure}
\begin{center}
\epsfxsize=10cm \epsfbox{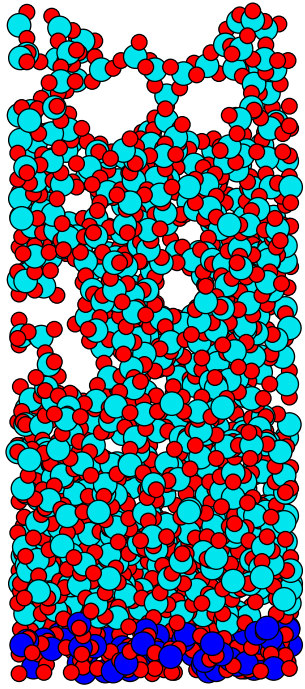}
\caption {
Snapshot of the sample grown with 1 eV clusters. The total number of atoms is
2040. Silicon atoms are in light grey while oxygen atoms are dark grey.
Darker atoms at the bottom of the cell indicate the fixed layers in the
substrate region.
}
\label{snap}
\end{center}
\end{figure}

\begin{figure}
\begin{center}
\epsfxsize=6cm \epsfbox{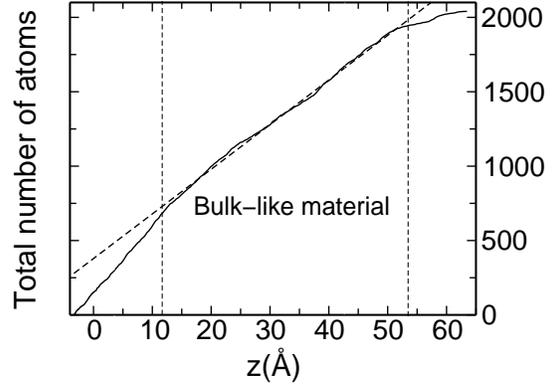}
\caption{Total (integrated) number of atoms as a function of the distance
from the bottom of the cell for the sample grown without bombardment ($R=0$).
The dashed curve is the best linear fit to the data in the bulk-like region of the
sample (as indicated).
}
\label{profdens}
\end{center}
\vspace{-0.5cm}
\end{figure}

In order to assess the role and importance of bombardment on the properties of
the grown material, we first discuss the case $R=0$, where all incoming
particles have the same energy, {\em viz.} 1 eV. Figure \ref{snap} shows a
snapshot of the system at the end of the run. The material is evidently highly
disordered and porous. In Fig.\ \ref{profdens}, we plot, for this system, the
corresponding integrated number of atoms from the bottom of the simulation cell;
we note that it increases approximately linearly with $z$ in the region 13-53
\AA. This portion will henceforth be referred to as the ``bulk-like'' part of
the deposit, where structural analysis will be performed in order to limit
artifacts due to the surface and the substrate. The same procedure is used
for samples grown with bombardment, which leaves us with 20 to 40 \AA\ of bulk
material, about 1000 atoms in all cases.

The overall mass density in the bulk-like region of the slab, 1.3$\pm$0.1
g/cm$^3$, is about 40\% less than the ``normal'' density of {\em a-}SiO$_2$ (2.2
g/cm$^3$) and 12\% less than the density of SiO$_2$ grown by electron-beam
evaporation on a silicon substrate at room temperature.\cite{brunet} The
discrepancy between experimental and calculated densities is possibly due to the
short duration of the simulation runs, which may not be long enough to allow
complete relaxation, and thus full densification. (The computational effort
required to achieve complete relaxation is formidable, and it constitues one of
the outstanding problems of disordered-system simulations). For this model of
silica, it is worth noting that the density is close to the critical value of
1.4 g/cm$^3$, below which pore percolation occurs and leads to
fracture.\cite{vash6} This was not observed here; however, as will be shown
below, the film contains high tensile stress ($\sim$ 2 GPa). Thus, relaxation is
expected to proceed by densification and/or fracture formation. This is
consistent with the observation of reduced crack toughness and adhesion to the
substrate for the case of coatings grown at low substrate temperature, $T_S$,
with insufficient bombardment.\cite{smidt}

\begin{figure}
\begin{center}
\vspace{-0.5cm}
\epsfxsize=10cm \epsfbox {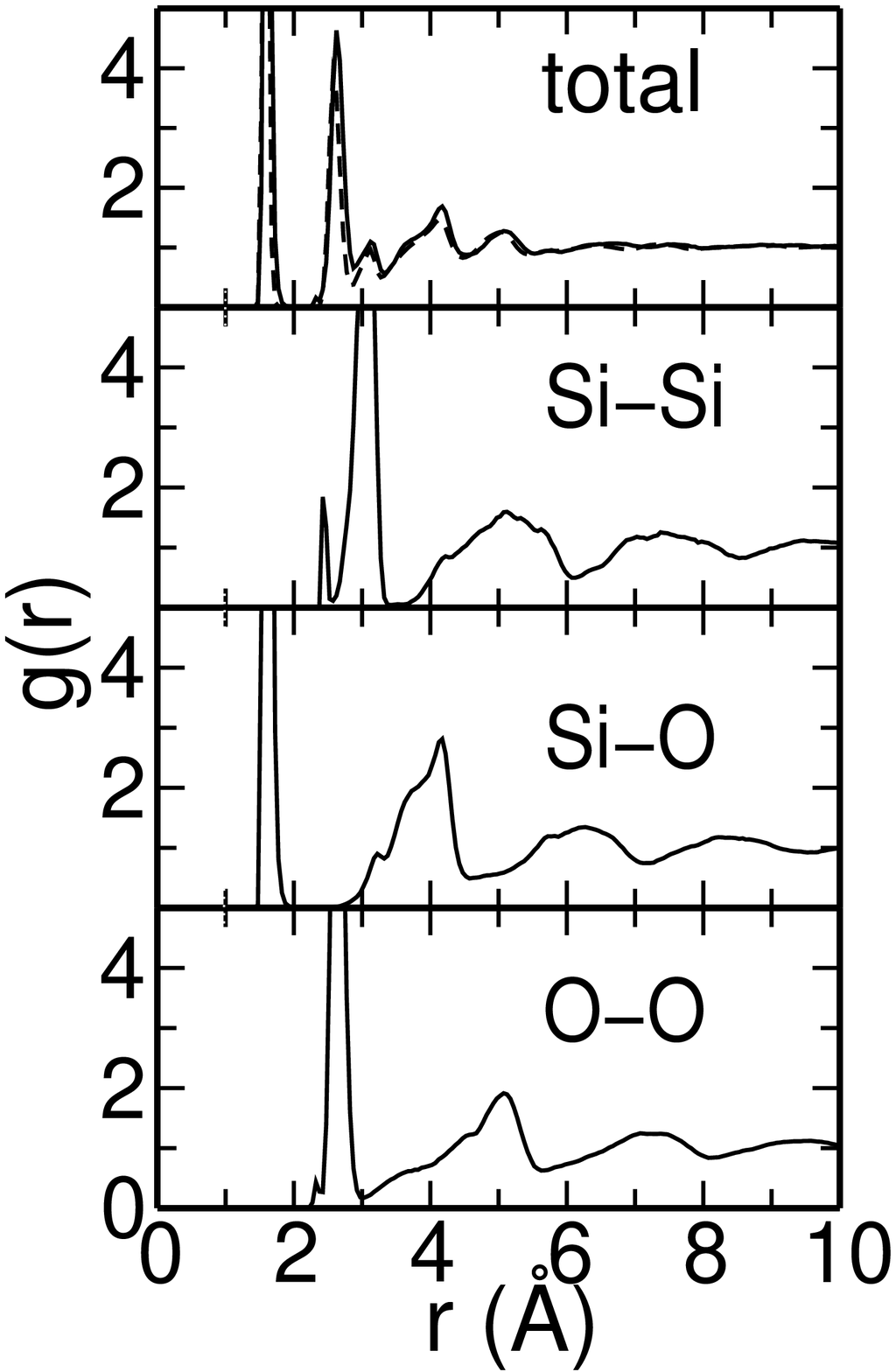}
\caption{ Total and partial pair distribution functions calculated for a
sample grown without bombardment ($R=0$). In the top panel, the dashed line
shows the total pair distribution function for bulk amorphous silica prepared
using ART.\protect\cite{mousseau}
}
\label{rdfpartial}
\end{center}
\end{figure}

We now discuss the structure and topology of the deposited material, in order
to provide a reference for samples grown in the presence of bombardment (cf.\
Sec.\ \ref{with}). Figure \ref{rdfpartial} shows the total pair distribution
function for the bulk-like part of the sample, as well as for bulk silica; also
shown are the partial distributions $g_{\rm SiSi}$, $g_{\rm SiO}$ and $g_{\rm
OO}$. Here again, the amorphous character of the material is clearly evident:
order beyond the nearest-neighbour shell rapidly fades out. The
nearest-neighbor Si--O, Si--Si and O--O distances are listed in Table
\ref{data}, along with those of bulk {\em a-}SiO$_2$,\cite{johnson} as well as
other material produced by secondary ion deposition (SID) at room
temperature\cite{george} and LPCVD at $T_S=430^\circ$ C; \cite{konnert} the
calculated and experimental values are seen to be in satisfactory agreement.
The partial pair correlations reveal a broadening of the first peak of the
Si--Si distribution compared to the ``normal'' glass, presumably a consequence
of the more disordered structure of the deposited film. Also, small pre-peaks
to the left of the nearest-neighbour peak are seen to develop in the Si--Si and
O--O correlation functions, at 2.45 and 2.32 \AA\ respectively. As will be
demonstrated below, these are due to the presence of two- and three-membered
rings, leading to Si--O--Si and O--Si--O bond angles of 90 and 130 degrees,
respectively.

\begin{figure}
\begin{center}
\vspace{-0.5cm}
\epsfxsize=10cm \epsfbox{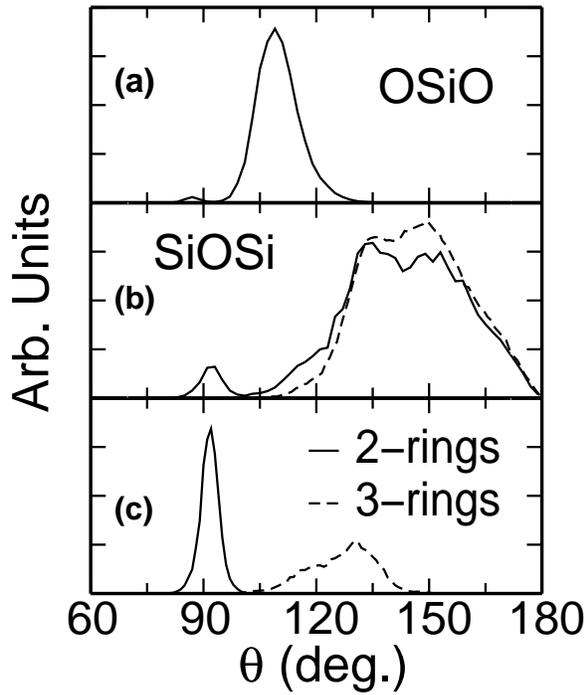}
\caption{Bond-angle distributions for a sample grown without bombardment: (a)
O--Si--O; (b) Si--O--Si; also shown is the corresponding distribution for
bulk-amorphous silica produced with ART (dashed line); (c) Si--O--Si for
the special cases of two- and three-membered rings.
}
\label{compare}
\end{center}
\end{figure}

In spite of the disordered nature of the network, the structural building
block remains, to a good approximation, the Si(O$_{1/2}$)$_{4}$ tetrahedron,
even though it might be slightly distorted. Indeed, the coordination of
silicon atoms, calculated from the area under the first peak of g$_{\rm
SiO}$, is about 3.9. Further, as displayed in Fig. \ref{compare}(a), the
O--Si--O bond angle distribution has a major peak centred at 109.7$^\circ$,
the usual tetrahedral bond angle.

The topology of the network can be described more precisely in terms of rings
of nearest-neighbor bonds, where an $n$-membered ring is defined as the
shortest closed path of $n$ Si--O bonds originating on a given silicon atom.
\cite{vash0} In normal silica glass, the statistics are dominated by five-,
six-, and seven-fold rings.\cite{vash5} Smaller rings are unlikely to occur
because they are too expensive in terms of elastic bond-bending energy. The
existence of ``odd-numbered'' rings results in a broad distribution of
Si--O--Si bond angles. In the normal material, this distribution is centered
around 143$^\circ$ and has a width of about 25$^\circ$,\cite{chan} as can be
seen in Fig.\ \ref{compare}(b). In contrast, the mean Si--O--Si bond angle for
evaporated SiO$_2$ is known from infrared spectroscopy measurements to be
7--10$^\circ$ smaller than in {\em a-}SiO$_2$.\cite{brunet} This is attributed
to the presence of three- and four-membered rings, manifested by the presence
of two sharp peaks at 606 and 495 cm$^{-1}$ in the Raman spectrum of chemical
vapor deposited (CVD) films, produced at low substrate temperature.\cite{devine}

Such small rings are a common feature of disordered SiO$_2$ networks, and they
are a consequence of rapid quenching rates;\cite{devine,mcmillan} thus, they are
also expected to be present in SiO$_2$ produced using evaporation methods. In
the present model, the fraction of three- and four-membered rings is about 0.1
and 0.15 per deposited silicon atom, respectively, roughly ten and two times
that of normal {\em a-}SiO$_2$.\cite{vash7} The mean value of the Si--O--Si bond
angle is 139$^\circ$. These results are consistent with the above-mentioned
experimental results for evaporated films. In addition, two-membered rings
(edge-sharing tetrahedra) are present here, but absent in the normal material.
As can be seen in Fig.\ \ref{compare}(b), the presence of such highly-strained
rings gives rise to a sharp peak at low angles, $\sim92^{\circ}$, in the
Si--O--Si bond angle distribution, while the distributions at high angles agree
within statistical accuracy. The $92^{\circ}$ peak is directly related to
two-membered rings, as can be appreciated from Fig.\ \ref{compare}(c), while
three-membered rings give rise to a shoulder at about 130$^{\circ}$ in the
bond-angle distribution.

\subsection{Structure of samples grown {\em with} bombardment}\label{with}

We now examine the structure of samples produced in the presence of
bombardment. It is expected that the energy provided by energetic particles
will have an ``annealing'' effect, thereby reducing the proportion of
high-energy structures --- mostly two- and three-fold rings.

\begin{figure}
\begin{center}
\epsfxsize=10cm \epsfbox{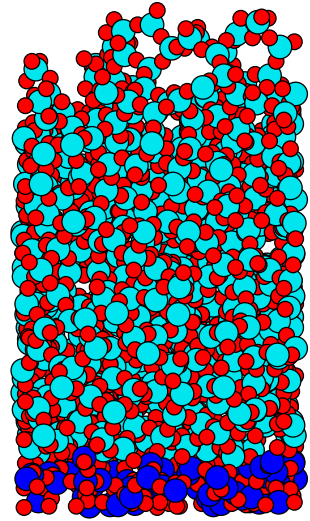}
\caption{
Snapshot of the samples grown with $R$=0.73. The total number of atoms is
2040. The effect of bombardment is clearly visible when comparing to
$R$=0.
}
\label{subsR07}
\end{center}
\vspace{-0.5cm}
\end{figure}

\begin{figure}
\begin{center}
\epsfxsize=8cm \epsfbox{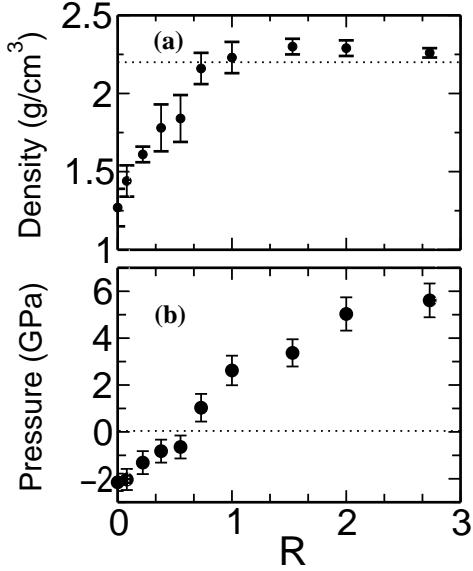}
\caption{ (a) Density and (b) pressure versus $R$ in the bulk part of the
various samples; in (a), the dotted line indicates the density of bulk silica
(2.2 g/cm$^{3}$); the error bars correspond to the calculated standard
deviations.
}
\label{densite}
\end{center}
\end{figure}

Simulations were carried out for several values of the $R$ parameter. As an
example, we show in Fig.\ \ref{subsR07} the final configuration for $R=0.73$.
Comparing this with Fig.\ \ref{snap}, the main role of bombardment ---
densification --- is clearly evident: the mass density of this new sample is
$\rho= 2.16 \pm$ 0.1 g/cm$^3$, much higher than the value of 1.3 g/cm$^3$ for
the sample grown without bombardment, and close to 2.2 g/cm$^3$ for normal
{\em a-}SiO$_2$. (The density of $\alpha$-quartz is 2.65 g/cm$^3$). In Fig.\
\ref{densite}(a), we plot $\rho$ as a function of $R$; it is seen to increase
rapidly, reaching a plateau at about 2.3 g/cm$^{3}$ for $R \geq 0.6 \equiv
R_c$. This ``critical'' $R$ value roughly marks the crossover from a state of
tensile stress (at small values of $R$) to a state of compressive stress, as
can be seen in Fig.\ \ref{densite}(b) where the pressure is plotted vs $R$.
Near $R_c$, the ratio of ion-to-condensing particle fluxes $\phi_i/\phi_n
\approx 0.4-0.5$; this yields a critical energy, $E_c$, for obtaining
high-density, low-stress films, $E_c=E_i \phi_i/\phi_n$, on the order of
12--15 eV per condensing SiO$_2$ unit.

The calculated value of $E_c$ is in reasonable agreement with experiment,
considering the relatively crude nature of our model and the evident difficulty
in measuring this quantity in a reliable manner. Indeed, values have been
reported between 10 and 100 eV.\cite{ross,smidt} Souche {\em et al.},
\cite{souche} for instance, gave 35 eV, based on optical index measurements on
SiO$_2$ prepared by oxygen IBAD at $T_S=250^\circ$C, for a beam energy $E_i=150$
eV. McNeil {\em et al.},\cite{mcneil} using a similar deposition process but
lower $E_i$, 30 eV, found the optical transmittance to deteriorate with high ion
current density, indicating that $E_c$ is {\em smaller} than 30 eV; an estimate
in the range 10--20 eV has been given.\cite{harpernimb} Finally, Al-Bayati {\em
et al.},\cite{albayati} using 10 eV Si$^+$/O$^+$ ion beams to grow SiO$_2$ on a
silicon substrate at 350$^\circ$C, found the interface to be smooth and free of
strain, and the film to be of excellent quality, with no evidence of suboxides;
thus, 10 eV would appear to be an upper bound to $E_c$.

The structural changes which take place upon densification can be investigated
in terms of short- and medium-range order modifications. At short range, Si--O
and O--O nearest-neighbour distances are found to depend very little on $R$, and
they are close to the values observed for the sample grown without bombardment
(cf.\ Section \ref{without}). We note, however, a slight ($\sim$2\%) increase of
the Si--Si distance, consistent with an increase of the Si--O--Si mean bond
angle from 139$^\circ$ for $R=0$ to 143$^\circ$ for $R=2.73$. The Si
coordination, between 3.90 and 3.96, shows no systematic variation with rising
density.

\begin{figure}
\begin{center}
\vspace*{-0.0cm}
\epsfxsize=7cm \epsfbox{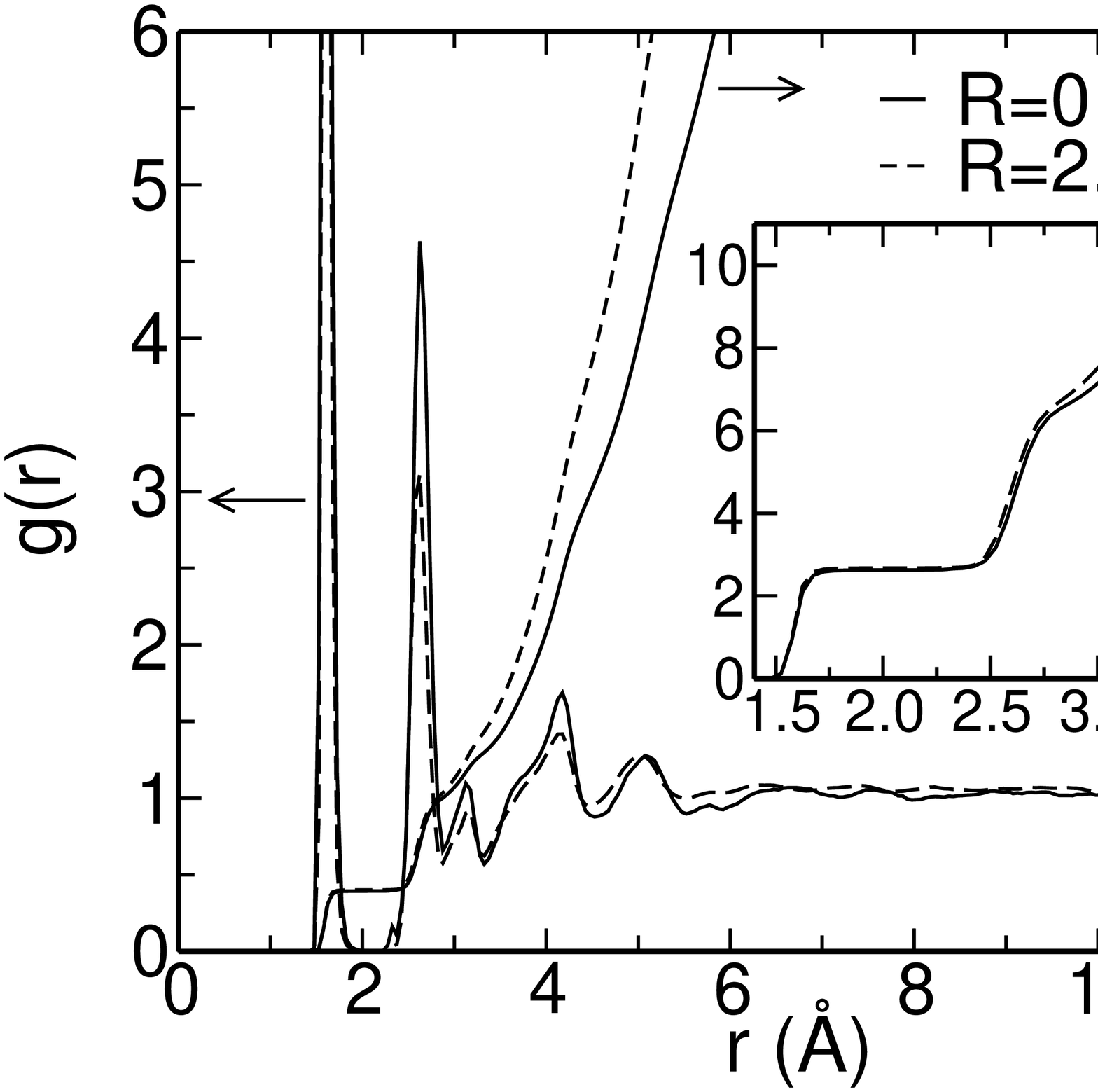}
\caption{
Total pair distribution function, $g(r)$, and running coordination number,
$N(r)$, for the samples grown with $R=0$ and $R=2.73$, as indicated. The
inset is a magnification of $N(r)$ in the region 1.4--3.5 \AA .
}
\label{rdfcomp}
\end{center}
\end{figure}

However, the picture is quite different at intermediate length scales. In
Fig.\ \ref{rdfcomp} we plot, for both $R=0$ and $R=2.73$, the total pair
distribution function, $g(r)$, and the running coordination number, $N(r)$;
the latter, essentially the integral of the former, gives the mean number of
atoms within a sphere of radius $r$ centered on an average atom, irrespective
of the atomic species. It is clear that the $N(r)$ curves for the two samples
are very similar for $r \leq 3$ \AA, confirming the role of the
Si(O$_{1/2}$)$_{4}$ tetrahedron as the structural building block (no
homoatomic bonds are allowed in this model). Above this value, however,
strong differences develop. Densification, therefore, involves structural
modifications at or beyond second nearest-neighbors.

\begin{figure}
\begin{center}
\vspace{-0.5cm}
\epsfxsize=7cm \epsfbox{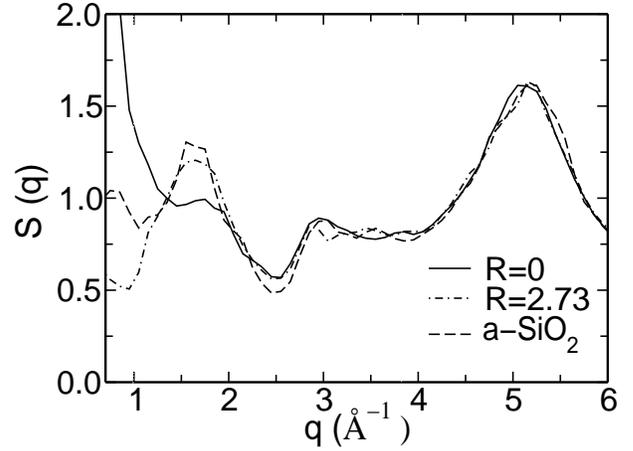}
\caption{
Total static structure factor for the samples with $R=0$, $R=2.73$, and for
amorphous silica. Data are truncated at 0.7 \AA$^{-1}$ due to the finite-size
of the simulation cell.
}
\label{ssf}
\vspace{-0.5cm}
\end{center}
\end{figure}

Amorphous silica is known to possess significant order at medium range,
manifested in the so-called ``first sharp diffraction peak'' (FSDP) at 1.55
\AA$^{-1}$, and observed in neutron and x-ray diffraction
experiments.\cite{moss} Although the precise origin of this peak, in terms of
stable entities, is still the matter of debate,\cite{moss,elliott,gaskell} MD
simulations by Nakano {\em et al.} have shown that its position and shape arise
from correlations between atoms at distances as large as 12 \AA.\cite{vash5} The
FSDP can therefore be viewed as a signature of the presence of MRO in {\em
a}-SiO$_2$. In order to assess this picture within the present approach, that
is, to quantify possible modifications in MRO upon densification, we calculated
the static structure factor, $S(q)$, which is also accessible experimentally.
This is shown in Fig.\ \ref{ssf} for the samples with $R=0$ and $R=2.73$, as
well as for {\em a-}SiO$_2$ prepared using the ART method;\cite{mousseau} the
curves were smoothed using Gaussian functions of width 0.1 \AA$^{-1}$ so as to
mimic thermal as well as experimental broadening. There is evidently very little
difference from one sample to another for $q$-values larger than 4 \AA$^{-1}$,
consistent with the existence of a structural unit common to all three networks.
In contrast, the low-$q$ portion of the spectra exhibit significant differences.
In particular, for $q \lesssim 1$ \AA$^{-1}$, the intensity varies strongly with
$R$, in a manner which is inversely related to the density.

While the finite size of the system, plus periodic boundary conditions,
artificially contribute to the intensity at such low-$q$ values, they cannot be
solely responsible for the observed variations. Since all three samples are
treated in the same manner and are similar in size, these variations must be
related to the presence of pores in the low-density samples, an effect which can
be understood in term of a Babinet argument --- the scattering by a pore is
similar to that from a cluster having comparable dimensions.

\begin{figure}
\begin{center}
\vspace{-0.0cm}
\epsfxsize=7cm \epsfbox{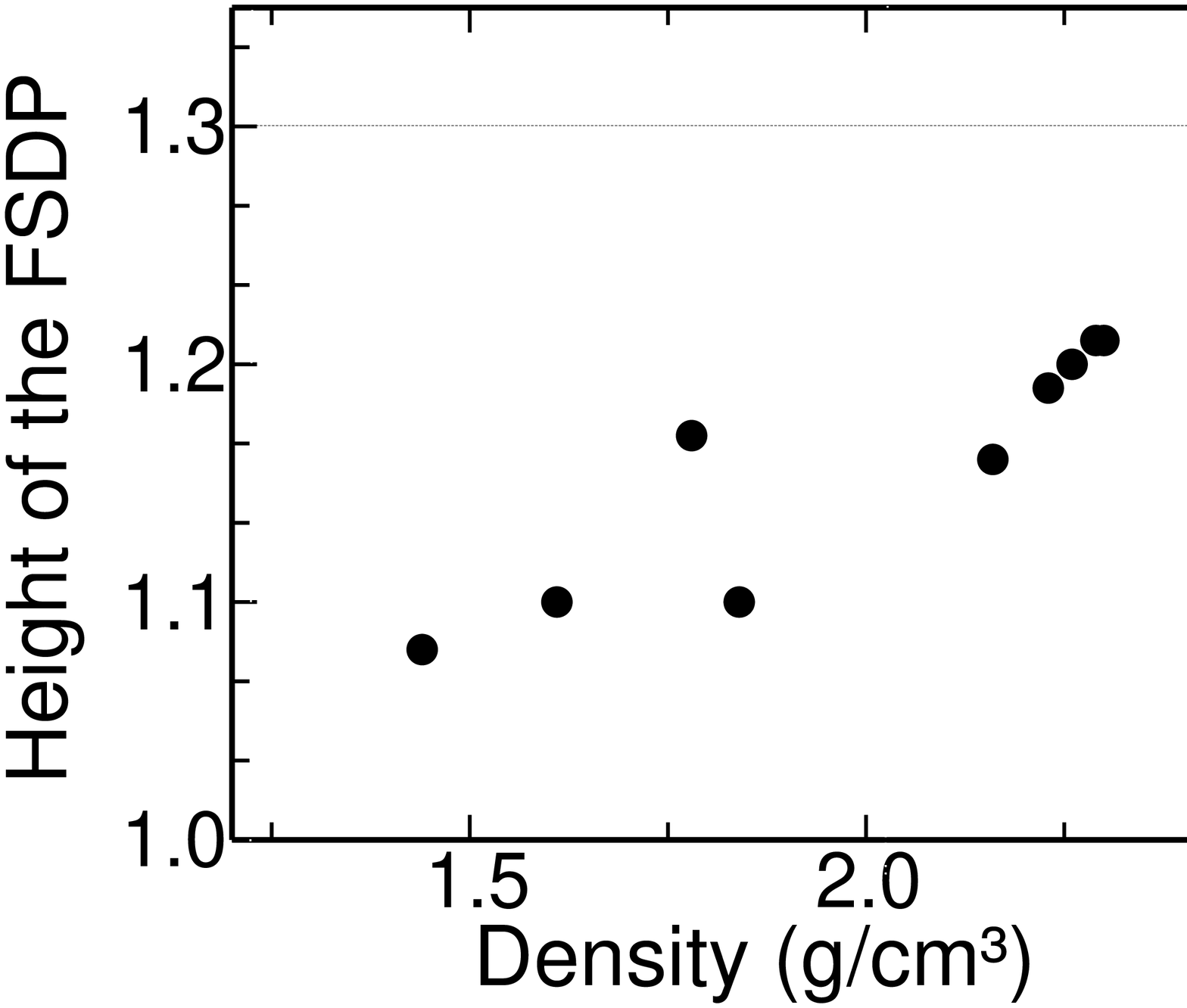}
\caption{
Height of the first sharp diffraction peak as a function of the density of
the samples. The dashed line is the corresponding value for {\em a-}SiO$_2$
prepared using ART.\protect\cite{mousseau}
}
\label{hfsdp}
\vspace{-0.5cm}
\end{center}
\end{figure}

Changes in the MRO upon densification are also apparent in the FSDP, which
the present model places at about 1.60 \AA$^{-1}$, in
satisfactory agreement with the experimental value, 1.55 \AA$^{-1}$. For
$R=0$, that is, highly porous material, the FSDP shows up as a mere shoulder,
whose intensity is 23\% smaller than that of {\em a}-SiO$_2$. For $R=2.73$, a
well-defined peak is observed, which is only 7\% smaller than that
of the normal material. The complete picture is provided in Fig.\
\ref{hfsdp}, where we plot the intensity of the FSDP as a function of
density. Although the statistics are not perfect, there is a clear tendency
for the height of the FSDP to increase with increasing density. It is
interesting to note that if the densification were the consequence of a
homogenous elastic compression of the network, that is, without structural
modifications, the mean atomic distances would decrease with increasing
density, implying an overall shift of $S(q)$ towards higher $q$ values. Here,
although the FSDP position is modified, no obvious trend is found.

The differences in heights of the FSDP among the various samples are
consistent with the results of Himmel {\em et al.},\cite{himmel} who found a
28\% smaller height in the x-ray static structure factor of electron-beam
evaporated silica, compared with bulk fused silica. Our calculations are also
in qualitative agreement with MD simulations of pressure-induced
densification of silica nanophases, where the height of the FSDP was found to
be 15 \% smaller compared to {\em a-}SiO$_2$ in the density range 1.67 to
2.03 g/cm$^3$.\cite{vash1}

\begin{figure}
\begin{center}
\vspace{-0.5cm}
\epsfxsize=8cm \epsfbox{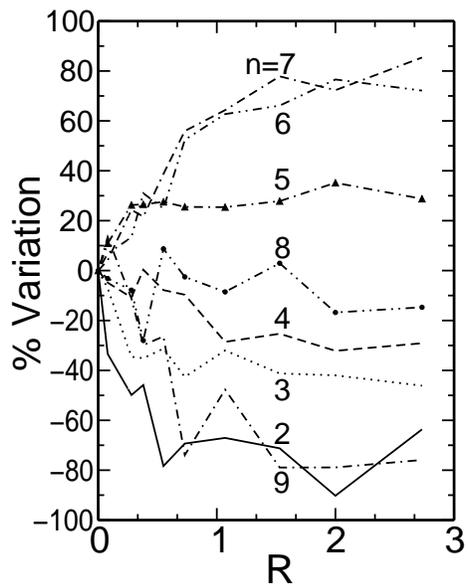}
\caption{Relative variation of the number of $n$-membered rings per silicon
atom as a function of R.
}
\label{anneau}
\vspace{-0.5cm}
\end{center}
\end{figure}

The ring structure of the various models can tell us more about changes in
MRO upon densification: because the size of rings is typically in the range
4--10 \AA,\cite{vash7} modifications in the MRO are related to those in the
arrangements of neighboring Si(O$_{1/2}$)$_{4}$ tetrahedra. The occurrence of
the various types of rings as a function of $R$, relative to $R=0$, is shown
in Fig.\ \ref{anneau}. In spite of statistical uncertainties, the following
clear trends can be identified. {\em (i)} The number of two-, three- and
four-membered rings decreases with rising $R$; indeed, as noted in Sec.\
\ref{without}, these correspond to high-energy structures and therefore are
expected to anneal out during low-energy bombardment. {\em (ii)} The number
of five-, six-, and seven-membered rings increases with rising $R$; these
rings being close to the most probable size in {\em a-}SiO$_2$, they are thus
strongly favored. {\em (iii)} The number of nine-membered rings decreases
with density, indicating a reduction in the number of pores. In
contrast, the relative occurrence of eight-membered rings exhibits no clear
trend, most likely because they are relatively stable entities in dense,
cristalline forms of silica, such as $\alpha$-quartz (2.65 g/cm$^3$) and
coesite (2.92 g/cm$^3$).\cite{liebau}

MD simulations on elemental silicon growth\cite{strickland,hensel} have
demonstrated, in agreement with experiments, that low-energy ion bombardment
has an effect similar to annealing. For the case of SiO$_2$ prepared by IBAD,
our results indicate that the elimination of two-, three- and four-fold rings
may be achieved with the use of a second particle beam with kinetic energy in
the range 12-15 eV. The concentration of three- and four-fold rings was
reported to decrease for CVD SiO$_2$ submitted to annealing;\cite{mcmillan}
however, to the best of our knowledge, no such investigation has been
reported regarding their dependence upon IBAD ion-current density or ion
kinetic energy. Such data would be of utmost interest.

The modifications in the ring statistics were pointed out in MD and
Monte-Carlo simulations of pressure-induced densification of bulk
silica.\cite{vash0,stixrude} It was found that most topological modifications
are mediated by defect annihilation and bond switching. The former mechanism
proceeds through the formation of bonds between non-bridging oxygen and
three-fold coordinated silicon atoms, while the latter involves the diffusion
of coordination defects (hopping of a dangling bond from one atom to a
neighbor, accompanied by the formation of a new bond).\cite{stixrude} This
has also been identified as an important relaxation mechanism in {\em
a-}SiO$_2$, and it appears to be characteristic of chemically-ordered
networks.\cite{mousseau} The latter point being an assumption of our model,
it is expected that these mechanisms also play an important role during
bombardment-induced densification. More precisely, since coordination numbers
do not show systematic variations with increasing density, bond-switching is
expected to dominate over the annihilation of coordination defects.

In pressure-densified material, the evolution of the ring populations is
quite different from the one we observe: the number of four-, eight-, and
nine-fold rings increase, while those of five- and six-fold rings, as
well as the intensity of the FSDP, decrease with increasing density (in the
range 2.2--3.2 g/cm$^3$). Taken together with the present results, this may
indicate that in its most stable form, {\em a-}SiO$_2$ has maximum MRO at
ambient temperature and pressure.

\section{Concluding remarks}\label{conclusion}

In this article, we have presented the results of extensive and detailed
molecular-dynamics simulations of the growth of SiO$_2$ on a {\em a-}SiO$_2$
substrate, with a view of understanding the process of densification induced
by low-energy particle bombardment. The physical picture which emerges from
our observations is the following: the network undergoes significant density
variations through structural changes at an intermediate length scale; more
precisely, ring statistics are affected, while the structural unit of the
network --- the Si(O$_{1/2}$)$_{4}$ tetrahedron --- remains essentially
unchanged. Upon bombardment, small rings, which are associated with defects
in the network, rearrange to produce larger ones, thus reducing the energy
cost associated with bond bending and increasing the mean Si--O--Si bond
angle. Large rings, particularly nine-membered ones, ``disintegrate'' into
five-, six-, and seven-membered rings, thus reducing the occupied volume and
causing the density to increase. Simultaneously, this influences the
intensity of the first sharp diffraction peak (FSDP), which is found to
increase with rising density. By calculating the pressure inside the
sample, we have found an ``optimal'' incident energy range for low-stress,
high-density films of 12 to 15 eV per condensing SiO$_2$ particle.

The very brief ($\sim$4 ns) real-time equivalent duration of our simulations may
not be sufficient to allow the samples to escape from local minima in the
potential energy surface, leading to structures which are incompletely relaxed
compared to the real material. Because relaxation processes, which are expected
to increase MRO correlations, may occur within the time scale of an experiment,
control over the film microstructure may be a difficult objective to attain when
performing ion-assisted deposition experiments. In addition, most of MRO
modifications arise from low-energy bombardment. For this reason, and as
demonstrated by experimental results,\cite{mcneil,albayati} we emphasize that
any deposition process where the average ion kinetic energy is greater than
about ten eV should be avoided. Besides ion-beam experiments, it was recently
demonstrated that magnetron sputtering processes could produce low-energy high
ion flux (up to 30 ions per neutral condensing particle), allowing amorphous
silicon films with a tunable amount of MRO to be grown.\cite{gerbi} We expect
that such findings will stimulate futher experiments on SiO$_2$ films, which
will allow our results to be assessed. For example, a systematic study of the
peaks associated with three- and four-membered rings in the Raman spectra of
silica coatings, and the evolution of the FSDP, with the kinetic energy and flux
of ions could help enhance our understanding of bombardment-induced
densification.

\vspace{0.5cm}

{\it Acknowledgments} -- A. L. is grateful to Gilles Dennler for useful
discussions. Normand Mousseau is acknowledged for providing the ART model. This
work is supported by grants from the Natural Sciences and Engineering Research
Council of Canada (NSERC) and by the ``Fonds pour la formation de chercheurs et
l'aide {\`a} la recherche'' (FCAR) of the Province of Qu{\'e}bec. We are
indebted to the ``R\'eseau qu\'eb\'ecois de calcul de haute performance''
(RQCHP) for generous allocations of computer resources.

\begin{center}
\begin{table}
\caption{Nearest-neighbor distances and Si--O--Si mean bond angle, calculated
for a sample grown without bombardment, along with those of SiO$_2$ produced
by secondary ion deposition (SID), electron beam evaporation, low-pressure
chemical vapor depostion (LPCVD), and bulk amorphous silica. Bond lengths are
given in \AA; in parentheses are the standard deviations.
}
\label{data}
\begin{tabular}{lcccc}
                                   & Si--Si       & Si--O        & O--O       & Si--O--Si       \\ \hline
This work                          & 3.0  (0.17)  & 1.61 (0.05)  & 2.64 (0.1) &  139$^\circ$    \\
SID a-SiO$_{2}$\cite{george}       & 3.06 (0.02)  & 1.61 (0.01)  & 2.64 (0.02)&  144.6$^\circ$  \\
e-beam a-SiO$_{2}$\cite{brunet}    &              &              &            &  136$^\circ$    \\
LPCVD a-SiO$_{2}$\cite{konnert}    & 3.08 (0.13)  & 1.61 (0.06)  & 2.60 (0.1) &                 \\
a-SiO$_{2}$\cite{johnson,chan}     & 3.08 (0.1)   & 1.61 (0.05)  & 2.63 (0.08)&  143$^\circ$    \\

\end{tabular}
\end{table}
\end{center}

\end{document}